\documentclass[graybox]{svmult}

 \pdfoutput=1
\usepackage{mathptmx}       % selects Times Roman as basic font
\usepackage{helvet}         % selects Helvetica as sans-serif font
\usepackage{courier}        % selects Courier as typewriter font
\usepackage{type1cm}        % activate if the above 3 fonts are
\usepackage{makeidx}         % allows index generation
\usepackage{graphicx}        % standard LaTeX graphics tool
\usepackage{multicol}        % used for the two-column index
\usepackage[bottom]{footmisc}% places footnotes at page bottom

\makeindex             % used for the subject index
\begin{document}

\title*{Enhancing gravitational wave astronomy with galaxy catalogues}
\author{Xilong Fan, Christopher Messenger, and Ik Siong Heng }
\institute{Xilong Fan \at School of Physics and Electronics Information, Hubei University of Education, 430205 Wuhan, China,\\
SUPA, School of Physics and Astronomy, University of Glasgow, Glasgow G12 8QQ, United Kingdom 
 \email{Xilong.Fan@glasgow.ac.uk}
\and Christopher Messenger \at SUPA, School of Physics and Astronomy, University of Glasgow, Glasgow G12 8QQ, United Kingdom  \email{Christopher.Messenger@glasgow.ac.uk}
\and Ik Siong Heng \at SUPA, School of Physics and Astronomy, University of Glasgow, Glasgow G12 8QQ, United Kingdom  \email{Ik.Heng@glasgow.ac.uk}}
\maketitle

\abstract*{Joint gravitational wave (GW) and electromagnetic (EM) observations, as  a key research  direction in \emph{multi-messenger astronomy}, 
will provide deep insight into the astrophysics of a vast range of astronomical phenomena. Uncertainties in the source sky location
estimate from gravitational wave observations mean follow-up observatories must scan large portions of
the sky for a potential companion signal.  A general frame of joint GW-EM observations  is presented  by a multi-messenger observational triangle. Using a Bayesian approach to multi-messenger astronomy, we investigate the use of
galaxy catalogue and host galaxy information to reduce the sky region over which follow-up observatories must scan, as well as study its use for improving the inclination angle estimates for coalescing binary compact objects.  We demonstrate our method
using a simulated neutron stars inspiral signal injected into simulated Advanced detectors noise and estimate
the injected signal sky location and inclination angle using the Gravitational Wave Galaxy Catalogue.   In this case study, the top three candidates in rank have $72\%$, $15\%$ and $8\%$ posterior probability of being the host galaxy, receptively.  The standard deviation of cosine inclination angle  (0.001) of the neutron stars binary  using gravitational wave-galaxy  information is much smaller than that (0.02) using only gravitational wave posterior samples.}

\abstract{Joint gravitational wave (GW) and electromagnetic (EM) observations, as  a key research  direction in \emph{multi-messenger astronomy}, 
will provide deep insight into the astrophysics of a vast range of astronomical phenomena. Uncertainties in the source sky location
estimate from gravitational wave observations mean follow-up observatories must scan large portions of
the sky for a potential companion signal.  A general frame of joint GW-EM observations  is presented  by a multi-messenger observational triangle. Using a Bayesian approach to multi-messenger astronomy, we investigate the use of
galaxy catalogue and host galaxy information to reduce the sky region over which follow-up observatories must scan, as well as study its use for improving the inclination angle estimates for coalescing binary compact objects.  We demonstrate our method
using a simulated neutron stars inspiral signal injected into simulated Advanced detectors noise and estimate
the injected signal sky location and inclination angle using the Gravitational Wave Galaxy Catalogue.   In this case study, the top three candidates in rank have $72\%$, $15\%$ and $8\%$ posterior probability of being the host galaxy, receptively.  The standard deviation of cosine inclination angle  (0.001) of the neutron stars binary  using gravitational wave-galaxy  information is much smaller than that (0.02) using only gravitational wave posterior samples.}

\section{Introduction} 
\label{sec:1} 
\begin{figure*}
  \begin{center}
    \includegraphics[width=\textwidth]{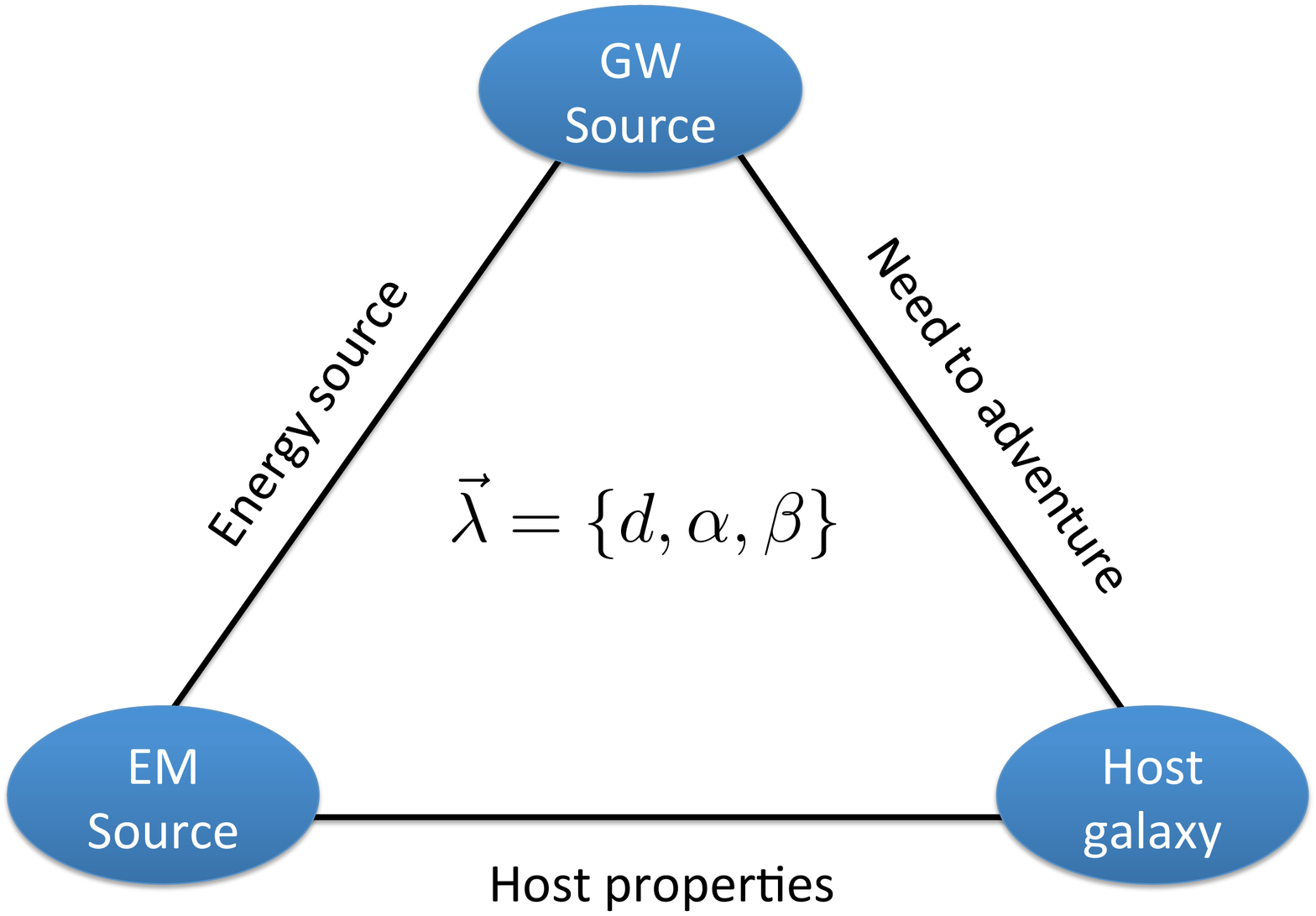}
    \caption{Multi-messenger observational triangle. The common parameters ($\vec{\lambda}$) of GW source, EM source and host galaxy include: distance ($d$), sky location ($\alpha, \beta$). Models and data have suggest the physical links between EM source and their host galaxies. The physical links between EM  and GW source have been investigated by theoretical   studies ,    while it is not clear for the link between  GW source and galaxies.  \label{ot}}
  \end{center}
\end{figure*}

The detection of Gravitational waves (GW) will herald a new era of astronomy, with Advanced LIGO~\cite{Harry:2010} and Advanced Virgo~\cite{Virgo:2009} expected to make the first detection of GWs within the next few years. \emph{Multi-messenger astronomy} involves the joint observation of
astrophysical phenomena by a combination of  GW, electromagnetic  (EM), and astroparticles observatories.  Examples of multi-messenger astronomy involving GWs include EM and  neutrinos observatories (e.g. \cite{2012ApJ...759...22K,Evans2012ApJS..203...28E,2013PhRvD..88l2004A,2013arXiv1303.2174B,2012A&A...539A.124L,AndoRevModPhys.85.1401,2014ApJS..211....7A,2012ApJ...760...12A,2010CQGra..27q3001A,2013ApJ...767..124N,2008CQGra..25r4034K}).  Galaxies, as the hosts of GW and/or EM sources, are examples of common observables in multi-messenger astronomy.  For example, searches for GWs in association with gamma-ray bursts (GRB) GRB070201 \cite{2008ApJ...681.1419A} and GRB051103 \cite{2012ApJ...755....2A} have ruled out the possibility that their progenitors are binary neutron stars (NS) sources in Andromeda galaxy (M31) and M81, respectively.   The basic assumption of this research is that there are common parameters among observations of GWs, GRBs and their host galaxy: sky location and distance.  Besides of these common parameters for GW sources, EM sources and host galaxy, there are physical links between each other depending on the nature of sources.  For example, merging NSs or NS-black hole systems and the collapse of supermassive stars, as potential sources of inspiral and burst GW, are thought to be the most likely progenitors for short and long GRBs respectively (e.g. \cite{1986ApJ...308L..43P,1989Natur.340..126E,2005Natur.437..845F,2006ARA&A..44..507W} ).  Nearby (z $<$1) long-GRBs are more likely to be observed in irregular galaxies or the outermost regions of spiral disks \cite{Fan2010A&A...521A..73F}, which are usually metal poor, while short-GRBs are observed in all types of galaxy (e.g. \cite{2013ApJ...769...56F}). This general picture is summarized by the \emph{multi-messenger astronomy observational triangle} shown in Fig.~\ref{ot}.

The broad sky location estimates ($>10$ deg$^2$,  e.g. \cite{Wen2008JPhCS.122a2038W,Wen2010PhRvD..81h2001W,Veitch2010PhRvD..81f2003V,schutz2011CQGra..28l5023S,2011CQGra..28j5021F,2014PhRvD..89h4060S}) from GW observations is a challenge for joint EM-GW observation.  In addition to exploiting common parameters, the link between GWs and galaxies is used to improve the efficiency of searching for an EM counterpart of a GW.  With the help of a galaxy catalog, Nuttall and Sutton (2010) \cite{2010PhRvD..82j2002N} proposed a ranking statistic to identify the most likely GW host galaxy based on galaxy distance and luminosity and the sky position error box.  The scan area needed by an EM followup team could be reduced to the regions of sky most likely associated with the GW host galaxy.
%By far,  the link is designed to GW compact binary coalescence (CBC)  event rate  in a galaxy in GW community.  
%%%I don't understand the sentence above
In past searches, blue band luminosity, which is assuming to be the tracer of the star formation rate of a galaxy, is used to estimated the rate of compact binary coalescences (CBC) (e.g. \cite{Evans2012ApJS..203...28E,2014ApJS..211....7A,2010PhRvD..82j2002N}).  Several other properties, such as morphology and metallicity, of galaxies are suggested to have effects on the CBC event rate in a galaxy via stellar population population synthesis studies (e.g. \cite{Belczynski2010ApJ...715L.138B, Fryer2012ApJ...749...91F,O'Shaughnessy2010ApJ...716..615O}).   Amongst the many existing galaxy catalogues, the Gravitational Wave Galaxy Catalogue (GWGC, \cite{White2011CQGra..28h5016W}) has been specifically compiled for current follow-up searches of optical counterparts from GW triggers.  GWGC is $\sim 100\%$ and $\sim 60\%$ complete out to about 40 and 100 Mpc respectively, estimated by blue band luminosity function (see discussion in \cite{White2011CQGra..28h5016W}).  Hanna et al. (2014) \cite{2014ApJ...784....8H} estimated that an average of $ \sim 500$ galaxies are located in a typical GW sky location error box for Advanced LIGO-Vigo network ($\sim$ 20 square degrees), up to range of 200 Mpc. The distance limitation of GWGC (or any other galaxy catalogs) comparing with  the GW detection horizon  leads  to  a  completeness issue for catalog  based multi-messenger astronomy studies.  An ongoing study by Messenger et al. (2014) \cite{chris2014} aims to address the  completeness issue.
 
Fan et al. (2014) \cite{Fan2014} have proposed a Bayesian approach to multi-messenger astronomy.  An example of  this approach is  the joint research of GW and it's galaxy host  with a  GW-galaxy relation  model (multi-messenger prior function in  \cite{Fan2014}).    One of merits of this Bayesian approach is that posterior probability of a galaxy hosting the GW source is estimated,  as well as  it's ranking in  host candidates.  The probabilities of GW host candidates can be used to guide EM follow-up observations to focus on the particular galaxy with very high posterior probability.  When the first few galaxies in the rank have similar posterior probabilities, the absolute rank becomes of less importance since these galaxies should be considered candidates of similar importance.  Moreover, better constraints on the distance of a galaxy-GW observations can benefit the inference on the inclination angle of a CBC event.   The better inclination angle  estimation could  help with  understanding the nature of  EM counterparts. For example, 
a set of inclination angles, combined with GRB observations, might constrain the beaming factor of GRBs, therefore the dynamics and the energetics of GRB jets and so on (e.g. \cite{2014arXiv1403.6917A}).

In this article, we present a case study of the Bayesian approach designed for joint EM and GW observations, in particular, the galaxy catalog (GWGC) and NS-NS coalescence events,  with a blue band luminosity based  multi-messenger prior function. The aim of this research is to identify the GW source host galaxy (see Sec.~\ref{sec2}) and  provide better inclination angle estimates (see Sec.~\ref{sec3}) .

\section{Identification of gravitational-wave host galaxies}\label{sec2}
\begin{figure*}
  \begin{center}
    \includegraphics[width=\textwidth]{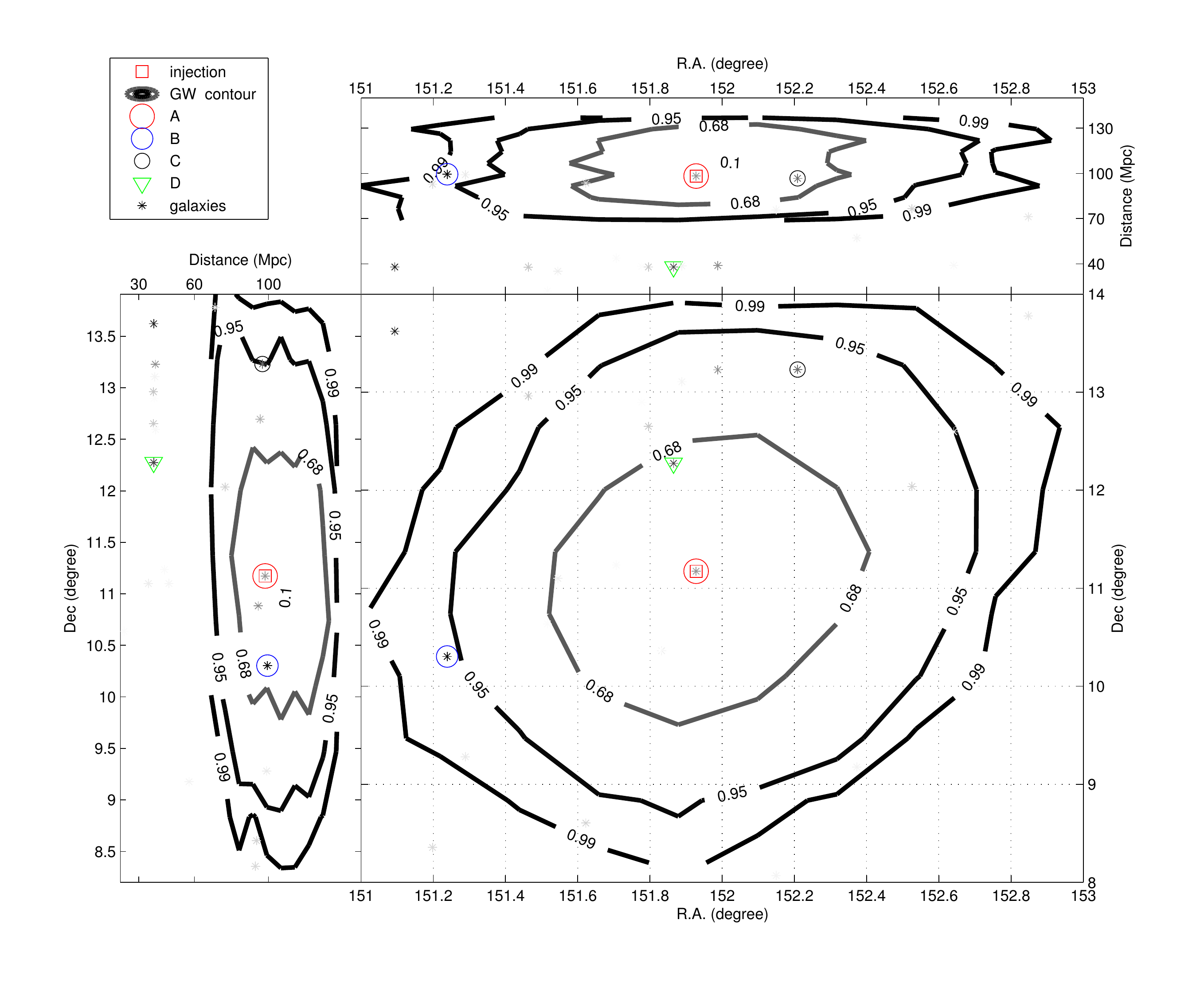}
    \caption{Sky localisation for a single BNS coalescence signal. The
      contours map out the 68\%, 95\% and  99\% confidence regions on the
      estimate of the sky location of the signal progenitor obtained
      using only GW observations. Also plotted are circle markers
      corresponding to the first (red circle), second (blue circle) and third
      (black circle) ranked host galaxy candidates, labelled A, B and C
      respectively, as determined using our Bayesian approach to multi-messenger astronomy.  The posterior  probability of A, B, C hosting the GW source  are  $72\%$, $15\%$  and  $ 8\%$, respectively.      Marker D (green downward-pointing triangle) has a lower probability because of its distance from the GW sky location estimate.    Additionally, grayscale
      asterisk markers are for all galaxies in this sky region,
      with the shade of the markers corresponding to each galaxies
      B-band luminosity. The darkest markers are the most luminous in
      the B-band. The BNS coalescence signal (red square) was injected at a
      sky location and distance  and in this case it
      corresponds to the top ranking galaxy candidate. The simulated
      signal  has an optimal network SNR 27.89\label{case_study_contours}}
  \end{center}
\end{figure*}

%Following the  Bayesian approach to multi-messenger astronomy built by \ref{}, 
We present a case study of simulated GW signals from NS-NS inspirals injected into simulated noise from the advanced LIGO-Virgo network.  The sky location of the injected NS-NS signal is randomly chosen from the locations of galaxies within the GWGC.
%according to a chosen astrophysical prior function based on blue band luminosity. The simulation detail can be  found in \ref{}. 
In Fig. \ref{case_study_contours}, we plot a 12 square degree region around the sky location where a simulated GW signal was injected. The galaxies in this region of the sky are plotted with grayscale asterisk markers. Bolder markers are galaxies with strong B-band luminosities. The top three galaxy candidates are marked by circles and the top galaxy marked by the largest circle is the injection. It is interesting to note that galaxy marked as D in Fig. \ref{case_study_contours}, which is within the 1 $\sigma$ skymap error area, is automatically excluded by the incompatible distance estimate provided by the GW signal analysis.  One of merits of this Bayesian approach is that the galaxy candidates are presented by ranking \emph{and} posterior probability.  In this case, the top three candidates in rank have $72\%$, $15\%$ and $8\%$ posterior probability of being the host galaxy, receptively. Therefore, EM follow-up teams could, with relative confidence, focus only on the top candidate in rank.  
%$p(\com,\EMnon|\EMdat,I)$.

\section{Enhanced inclination angle inference}\label{sec3}
Once the top ranked galaxies are identified, the distance of the candidates can be used to provide an improved estimate on the inclination angle ($\iota$) of the CBC progenitor. For example, each candidate in the GW signal error region is assigned an estimate of the posterior probability of hosting the GW signal.  Only the GW posterior samples within certain distance ranges should have non-zero probability. Galaxies with non-zero probability tend to be at the center of a distance range with $10\%$ of that distance as the radius of that range. Therefore, the GW posterior distribution combined with  the probabilities  of  the galaxy candidates  hosting the GW source  at  certain distances  can lead to a better estimation on inclination angle.

\begin{figure}%[h!]
\begin{center}
\includegraphics[width=\textwidth]{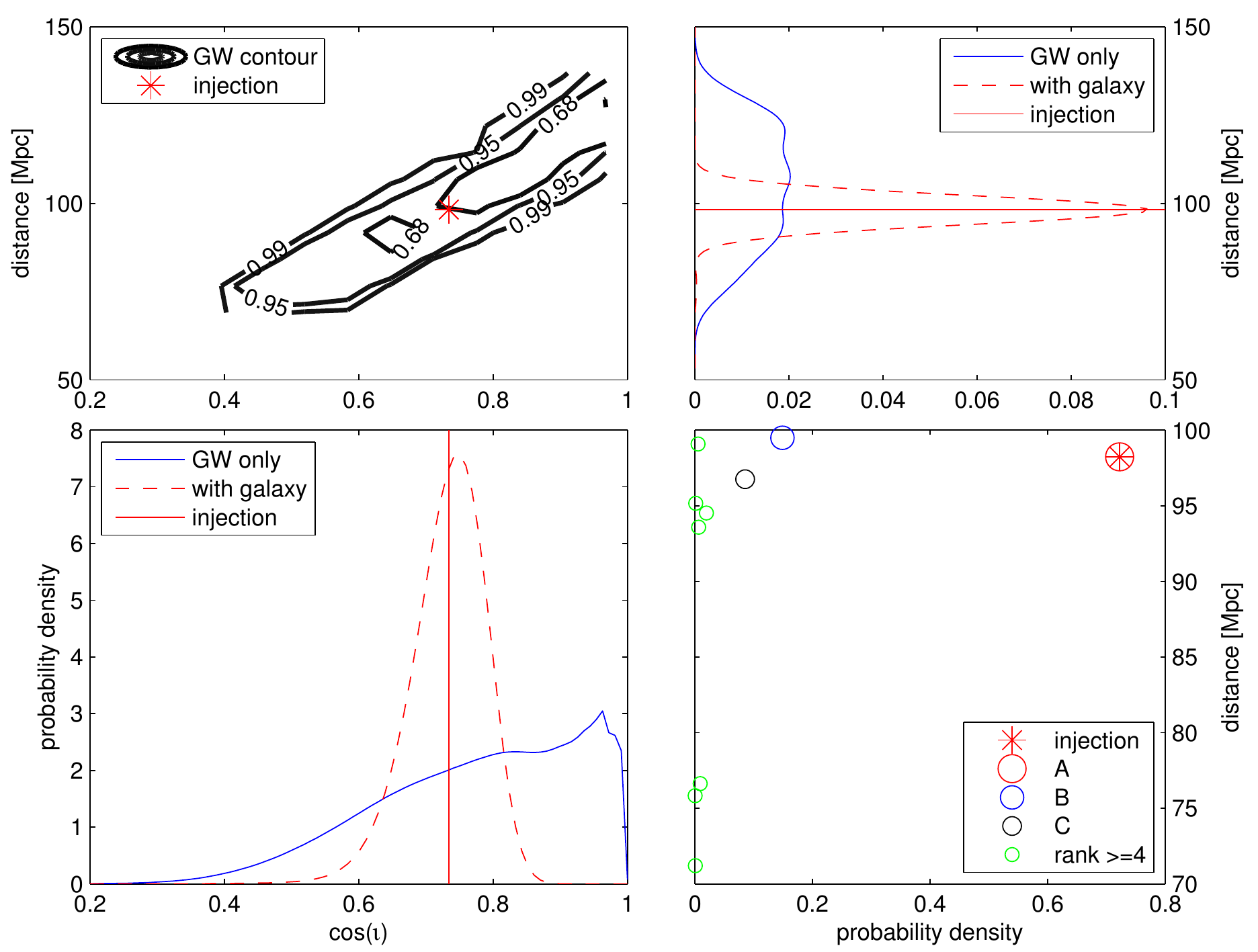}%colormap_gwgc_adv_allfree_11_lb} % color_map_ra_dec_gwgc_2}
\caption{An example showing the reduction in the degeneracy between distance and inclination angle
  $\iota$ .   Top-left panel correspond to GW posterior samples contour. Bottom-right  panel  correspond to posterior  probability of  host galaxy, and  plotted are circle markers corresponding to the first (red), second (blue) and third (black) ranked host galaxy
  candidates, labelled A, B and C respectively, as determined using our Bayesian approach to multi-messenger astronomy.  Top-right and bottom-left  panels correspond to probability density function of distance  and $\cos(\iota)$ estimated by Kernel smoothing function estimate,  receptively. Blue lines  and red dished lines correspond to  the  probability density function using only GW posterior samples and GW-galaxy information, receptively.    Injection value of distance and $\cos(\iota)$ are shown in red lines.  Using GW-galaxy  information, the standard deviation of $\cos(\iota)$  (0.001) is much smaller than that (0.02) using only GW posterior samples.  The simulated signal is the same one used in Fig~\ref{case_study_contours}.  \label{case_study_iota1}}
\end{center}
\end{figure}

In Fig.~\ref{case_study_iota1}, for the same simulation shown in Sec.~\ref{sec2},  we see that the posterior probability peaks strongly around the correct value of inclination. Furthermore, the inclination angle ($\iota$) posterior distribution is now much narrower given the additional distance information provided by the galaxy  candidates. In this example, the standard deviation of cos$(\iota)$ (0.001) obtained from joint GW-galaxy information is much smaller than that (0.02) obtained via a GW analysis alone. It is worth noting that the host galaxy for the GW signal does not need to be the top ranked galaxy for the inclination angle inference to improve. Even the true host galaxy is not be identified due to a cluster of galaxies within the GW location estimate, the additional distance information provided by all galaxies within the GW location estimate will still reduce the width of the inclination angle posterior. 
\section{Conclusions}
The implication of a Bayesian approach to multi-messenger astronomy is presented using a  NS-NS inspiral GW signal injected into simulated advanced detectors noise assuming sky and distance information drawn from the GWGC. A merit of this approach is that  both the rank and the posterior probability of a galaxy hosting the GW source are estimated with the help of a GW-galaxy relation model (blue band luminosity based multi-messenger prior function in this case study).  With this information, EM follow-up observation could focus on the first galaxy in the rank with a very high probability.  Once the host galaxy is identified, the additional distance information could benefit the inference of the inclination angle of CBC, shown in our example as the reduction of the variance in the estimation on inclination angle.   In this case study,  the posterior probabilities of  the top three candidates being the host galaxy in rank are $72\%$, $15\%$ and $8\%$, receptively.  The standard deviation of cosine inclination angle  (0.001) using GW-galaxy  information is much smaller than that (0.02) using only GW posterior samples.  The results reported in this research depend on various selection effects, such as the  dependence of the EM data on the common parameter set ($\vec{\lambda}$) (e.g. multi-messenger prior function, see discussion in \cite{Fan2014}), the  completeness of the EM data.  The limited range of the GWGC with respect to the expect range for Advanced detectors highlights the need to take into account selection effects introduced by the catalogue and elsewhere.   We will address the  completeness issue in next work \cite{chris2014}.

\begin{acknowledgement}
We would like to acknowledge valuable input from J.~Kanner. The authors gratefully acknowledge the support of this research by the
Royal Society, the Scottish Funding Council, the Scottish Universities
Physics Alliance and and the Science and Technology Facilities Council
of theUnited Kingdom.  XF acknowledges financial support from National Natural Science Foundation of China 
(grant No.~11303009).  XF is a Newton Fellow supported by the Royal
Society and CM is a Lord Kelvin Adam Smith Fellow supported by the
University of Glasgow.\end{acknowledgement}

%
%\section*{Appendix}
%\addcontentsline{toc}{section}{Appendix}
%
%
%When placed at the end of a chapter or contribution (as opposed to at the end of the book), the numbering of tables, figures, and equations in the appendix section continues on from that in the main text. Hence please \textit{do not} use the \verb|appendix| command when writing an appendix at the end of your chapter or contribution. If there is only one the appendix is designated ``Appendix'', or ``Appendix 1'', or ``Appendix 2'', etc. if there is more than one.

%%%%%%%%%%%%%%%%%%%%%%%% referenc.tex %%%%%%%%%%%%%%%%%%%%%%%%%%%%%%
% sample references
% %
% Use this file as a template for your own input.
%
%%%%%%%%%%%%%%%%%%%%%%%% Springer-Verlag %%%%%%%%%%%%%%%%%%%%%%%%%%
%
% BibTeX users please use
% \bibliographystyle{}
% \bibliography{}
%
\bibliographystyle{spphys.bst}
\bibliography{Bibliography}

\end{document}